\documentclass[12pt]{article} 
\usepackage{a4}
\usepackage{epsfig}
\usepackage{a4wide}
\usepackage{wasysym}
\def\Jo#1#2#3#4{{#1} {\bf #2}, #3 (#4)}

\def\NPB{{Nucl. Phys.} {\bf B}}
\def\NPBP{{Nucl. Phys.} {\bf B} (Proc. Suppl.)}
\def\PLB{{Phys. Lett.}  {\bf B}}
\def\PRL{Phys. Rev. Lett.}
\def\PRD{{Phys. Rev.} {\bf D}}
\def\EPC{{Eur. Phys. J.} {\bf C}}
\def\EPA{{Eur. Phys. J.} {\bf A}}
\def\IJMP{Int. J. of Mod. Phys. {\bf A}}
\def\JHEP{JHEP}
\def\PTP{Prog. Theo. Phys.}
\def\MPLA{{Mod. Phys. Lett} {\bf A}}
\def\PPNP{Prog. Part. Nucl. Phys.}
\def\NIMA{{Nucl. Instrum. Methods} A}
\def\NPPS{Nucl. Phys. Proc. Suppl.}

\def\ra{\rightarrow}
\def\be{\begin{equation}}
\def\ee{\end{equation}}
\def\gs{\mathrel{
   \rlap{\raise 0.511ex \hbox{$>$}}{\lower 0.511ex \hbox{$\sim$}}}}
\def\ls{\mathrel{
   \rlap{\raise 0.511ex \hbox{$<$}}{\lower 0.511ex \hbox{$\sim$}}}}
\newcommand{\obb}{0\mbox{$\nu\beta\beta$}}
\newcommand{\onbb}{neutrinoless double beta decay }

\newcommand{\ba}{\begin{array}{c}}
\newcommand{\baz}{\begin{array}{cc}}
\newcommand{\bad}{\begin{array}{ccc}}
\newcommand{\bea}{\begin{equation} \begin{array}{c}}
\newcommand{\eea}{ \end{array} \end{equation}}
\newcommand{\ea}{\end{array}}
\newcommand{\D}{\displaystyle}
\newcommand{\dms}{\mbox{$\Delta m^2_{\odot}$ }}
\newcommand{\dma}{\mbox{$\Delta m^2_{\rm A}$ }}
\newcommand{\meff}{\mbox{$\langle m \rangle$}}

\newcommand{\tm}{\mbox{$\tilde{m}$}}
\def\ra{\rightarrow}
\newcommand{\ppp}{\mbox{$(+++)$ }}
\newcommand{\pmm}{\mbox{$(+--)$ }}
\newcommand{\mpm}{\mbox{$(-+-)$ }}
\newcommand{\mmp}{\mbox{$(--+)$ }}

\begin{document}
\newpage
\title{\hfill { {\small DO--TH 02/01}}\\
\hfill { {\small hep-ph/0203214}}\\ \vskip 1cm  
{\bf Measuring leptonic $CP$ Violation in Neutrinoless Double Beta Decay}}
{\it 
\author{Werner Rodejohann\footnote{Email address: 
rodejoha@xena.physik.uni-dortmund.de}\\
{\normalsize Lehrstuhl f\"ur Theoretische Physik III,}\\ 
{\normalsize Universit\"at Dortmund, Otto--Hahn--Str. 4,}\\ 
{\normalsize 44221 Dortmund, Germany}} 
}
\date{}
\maketitle
\thispagestyle{empty}
\begin{abstract}
We investigate under which circumstances one can show 
the existence of leptonic $CP$ violation with the help of a 
positive or negative signal in neutrinoless double beta decay. 
The possibilities of cancellations are investigated for 
special mass hierarchies and the different solar solutions. 
The possibility that the mixing angle connected with the solar 
neutrino problem is smaller or larger than $\pi/4$ is taken into 
account. The non--maximality of that angle in case of the LMA solution 
allows to make several useful statements. 
The four different $CP$ conserving 
possibilities are analyzed. 
It is implemented how precisely the oscillation parameters will be known 
after current and future experiments have taken data. 
The area in parameter space, in which $CP$ 
violation has to take place, is largest for the LOW solution and in general 
larger for the inverse mass scheme. 
  
\end{abstract}

\newpage
\section{\label{sec:eins}Introduction}
Evidence for lepton flavor violation has been collected 
in large amounts, courtesy of neutrino oscillation 
experiments \cite{nuexp,SNOnew}. An explanation of the smallness of the 
implied neutrino masses is given by the see--saw mechanism \cite{seesaw}, 
which introduces Majorana neutrinos and thus lepton number violation 
to the theory. In recent years, the search for this phenomenon has 
concentrated on \onbb (\obb). The decay width of this process 
is proportional to the square of the so--called effective mass 
of the electron neutrino, 
\be \label{eq:meff}
\meff = \sum U_{ei}^2 m_i \; , 
\ee
where $U$ is the leptonic Maki--Nakagawa--Sakata (MNS) mixing 
matrix \cite{MNS}. 
Since \meff{} depends on the neutrino masses, the two mixing angles 
connected with solar and reactor experiments and two phases in $U$, 
any measurement or non--measurement of \obb{} can in principle answer some 
of the open questions of neutrino physics, a topic which in the past 
has been addressed by a number of authors 
\cite{others0vbb,viss,petwol,petpas}. 
For example, by combining 
oscillation experiments and \obb, one can investigate 
the solution of the solar neutrino problem, 
the mass scheme, the value of the smallest mass eigenstate or 
the presence of leptonic $CP$ violation. 
In this note we 
shall concentrate on the latter point. For Majorana neutrinos there 
are three phases in $U$, two of which can in principle be measured 
through \obb.  
These two additional phases 
\cite{scheval1} are parameters of an extended Standard Model (SM) 
and are of interest e.g.\ regarding the stability of the neutrino mass 
matrix under radiative corrections \cite{stab} 
or in governing the magnitude of the baryon asymmetry 
of the universe via the leptogenesis mechanism \cite{leptogenesis}. 
Regarding the latter 
it has recently been shown in two quite different models 
\cite{wir,branco} that even for vanishing $CP$ 
violation in oscillation experiments, there can still be a sufficient 
baryon asymmetry generated. The amount of $CP$ violation found in 
\onbb is then crucial in testing different leptogenesis models. 
In contrast to 
the quark sector, there are four $CP$ conserving possibilities for 
Majorana neutrinos, all of which have different aspects. 
We shall discuss them in some detail, finding that in many cases 
they can be classified into two groups, sometimes even one single 
case can be identified. 
We decided to ignore the 
recently announced controversial indication for \obb{} \cite{evid}. 
See \cite{viss,comment} for a criticism of the statistical 
methods used in that analysis and \cite{reply} for replies. 
We shall only quote the measurement of the 
life--time limit on the \onbb of $^{76}$Ge, which is 
$1.5 \cdot 10^{25}$ y \cite{hemo}. Using different calculations 
for the nuclear matrix elements (NME), a limit on the effective mass 
of 
\be \label{eq:mefflimit}
\meff < (0.30 \ldots 0.97) \; \rm eV 
\ee
can be set. See \cite{NME} for a discussion of the different calculations. 
As future limits are concerned, several proposals for new 
experiments exist, such as 
CUORE \cite{CUORE}, EXO \cite{EXO}, MOON \cite{MOON}, 
Majorana \cite{MAJORANA}, or GENIUS \cite{GENIUS}, see 
\cite{rev} for a recent overview.  
As possible landmark limits we will assume 0.01 and $2 \cdot 10^{-3}$ eV, 
where the latter corresponds to the 10t GENIUS project. 
The expected uncertainty of the result is estimated to be 
around 20 to 30 $\%$ \cite{rev}.\\

The paper is organized as follows: in Section \ref{sec:form} the 
required formalism is briefly reviewed and in Section \ref{sec:CPC} the 
results are presented for some special mass hierarchies and 
then in Section \ref{sec:gen} 
the general case is analyzed, including the current and future 
uncertainty in the knowledge of the 
oscillation parameters. 
Section \ref{sec:sum} summarizes our conclusions.

\section{\label{sec:form}Formalism}
Since \meff{} is the absolute value of a sum of three complex numbers, 
it depends on two phases. 
The neutrino mixing matrix $U$ can be parameterized in a very 
convenient form, which treats these two phases as the 
two additional Majorana phases: 
\bea \label{eq:Upara}
U = U_{\rm CKM} \; 
{\rm diag}(1, e^{i \alpha}, e^{i (\beta + \delta)}) \\[0.3cm]
= \left( \bad 
c_1 c_3 & s_1 c_3 & s_3 e^{-i \delta} \\[0.2cm] 
-s_1 c_2 - c_1 s_2 s_3 e^{i \delta} 
& c_1 c_2 - s_1 s_2 s_3 e^{i \delta} 
& s_2 c_3 \\[0.2cm] 
s_1 s_2 - c_1 c_2 s_3 e^{i \delta} & 
- c_1 s_2 - s_1 c_2 s_3 e^{i \delta} 
& c_2 c_3\\ 
               \ea   \right) 
 {\rm diag}(1, e^{i \alpha}, e^{i (\beta + \delta)}) \, , 
\eea
where $c_i = \cos\theta_i$ and $s_i = \sin\theta_i$. 
Within the parameterization (\ref{eq:Upara}), \meff{} depends on $\alpha$ 
and $\beta$. The third phase $\delta$ can be probed in oscillation 
experiments \cite{osccpv}. 
For $CP$ 
conservation, different relative signs of the masses $m_i$ are 
possible, corresponding to the intrinsic $CP$ parities of the 
neutrinos \cite{phaalt,ichNPB}. Four situations are possible, 
with $m_i = \eta_i |m_i|$ one can write these cases 
as \ppp$\!\!$, \pmm$\!\!$, \mpm and \mmp$\!\!$, where the 
$(\pm\pm\pm)$ correspond to the relative signs of the mass states. 
Special values of the phases correspond to these 
sign signatures \cite{ichNPB}, 
\be \label{eq:CPV}
\bad (+++) & 
\eta_1 = \eta_2 = \eta_3 = 1 
& \leftrightarrow \alpha = \beta = \pi \\[0.2cm]
     (+--) & 
\eta_1 = - \eta_2 = - \eta_3 = 1 
& \leftrightarrow \alpha = \beta = \frac{\pi}{2}\\[0.2cm]
     (-+-) & 
\eta_1 = - \eta_2 = \eta_3 = -1 
& \leftrightarrow \alpha = \frac{\beta}{2} = \frac{\pi}{2}\\[0.2cm]
     (--+) & 
\eta_1 = \eta_2 = - \eta_3 = -1 
& \leftrightarrow \alpha = 2 \beta = \pi
\ea \, .
\ee
The neutrino mass itself is bounded by the result from the tritium spectrum 
\cite{me}, 
\be \label{eq:me} 
m_0 = \sum |U_{ei}|^2 m_i < 2.2 \, \rm eV \, . 
\ee 
Next generation laboratory \cite{KATRIN} as well as satellite 
\cite{PLANCK} experiments will be able to probe neutrino masses 
down to $\sim$0.4 eV\@. 
Finally, two possible mass schemes exist, the normal and inverse scheme 
can be written as follows: 
\be \label{eq:noin}
\bad \mbox{ \bf NORMAL} & & \mbox{ \bf INVERSE}\\[0.5cm]
m_3 = \sqrt{\dma + m_1^2} & & m_1 = \sqrt{\dma + m_3^2} \\[0.3cm]
m_2 = \sqrt{\dms + m_1^2} & & m_2 = \sqrt{-\dms + m_1^2} \; \; . \\[0.3cm]
m_1 = 0 \ldots 2.2 \mbox{ eV} & & m_3 = 0 \ldots 2.2 \mbox{ eV} \\
\ea  
\ee
Regarding the mixing angles and the mass squared differences, the 
best--fit points of the atmospheric oscillation parameters are \cite{atmBF}
\be \label{eq:atmBF}
\baz
\dma \!\! = 2.5 \cdot 10^{-3} \; {\rm eV^2}, & \tan^2 \theta_2 = 1 
\ea \, . 
\ee
The maximal \dma is about $5 \cdot 10^{-3}$ eV$^2$. 
As far as the solar solution is concerned, the Large Mixing Angle (LMA) 
solution is favored over the low \dms (LOW) solution 
by the latest data, especially after the recent 
SNO results \cite{SNOnew}.   
The other possibilities are the vacuum solution (VAC) and the 
Small Mixing Angle (SMA) solution, which are 
currently highly disfavored. We will therefore concentrate mainly 
on LMA and LOW\@. 
Typical best--fit points are \cite{solBF} 
\be \label{eq:solBF}
\bad
\dms \!\! = 4.5 \cdot 10^{-5} \; {\rm eV^2}, & \tan^2 \theta_1 = 0.4 
 & {\rm LMA} \\[0.4cm]
\dms \!\! = 1.0 \cdot 10^{-7} \; {\rm eV^2}, & \tan^2 \theta_1 = 0.7 
& {\rm LOW} \\[0.4cm]
\dms \!\! = 4.6 \cdot 10^{-10} \; {\rm eV^2}, & \tan^2 \theta_1 = 2.4 
& {\rm VAC} \\
\ea \; . 
\ee
The allowed ranges at 95 $\%$ C.L. of $t_1^2 \equiv \tan_1^2$ go from 
about 0.2 to 0.8 for LMA and 
0.5 to 1 for LOW\@. In fact, the upper value of $t_1^2$ for LMA 
differs in the different available analyses, which appeared after 
the new SNO results \cite{diffan}. 
We shall take 0.8 as an illustrative example. 
In case of the VAC solution, one may assume that 
$t_1^2$ lies between 0.4 and 3, i.e.\ also on the ``dark side'' of 
the parameter space. 
One may however safely state that maximal mixing is 
disallowed for the LMA solution, which will later on allow us to make 
some useful statements. See \cite{petnew} for a detailed 
analysis of the consequences of this fact. 
The mass scale \dms is lower than about $3 \cdot 10^{-4}$ eV$^2$ for 
LMA and $3 \cdot 10^{-7}$ ($3 \cdot 10^{-11}$) eV$^2$ for LOW (VAC). 
The third angle $\theta_3$ is constrained to be \cite{petbil} 
\be
\tan^2 \theta_3 \equiv t_3^2 \ls 0.08 \, . 
\ee
In the future, it will be possible to reduce the error of the atmospheric 
parameters \dma and $\sin^2 2 \theta_2$  
to about $10 \%$ by the MINOS, ICARUS and OPERA experiments 
\cite{barfut}. The SNO, KAMLAND and BOREXINO collaborations are 
able to reduce the errors 
on $\sin^2 2 \theta_1$ to about $5 \%$ and \dms 
will be known to a precision of a few $\%$ for LMA, about 
$10 \%$ for LOW and a few \permil for VAC \cite{visfut}. Long baseline 
experiments can make precision measurements of the atmospheric 
parameters on the few $\%$ level \cite{atmconv}, neutrino factories 
even at a 1 $\%$ level \cite{atmnufac}. Regarding $\tan \theta_3^2$, 
the experiments 
for the atmospheric scale as well as future reactor experiments 
\cite{futrea} can probe values of few$\cdot 10^{-3}$, 
long baseline experiments values of few$\cdot 10^{-5}$.

\section{\label{sec:CPC}$CP$ violation in special hierarchies}
\subsection{\label{sec:hie}Hierarchical scheme}
This scheme is realized for $m_3 \gg m_2 \gg m_1$. 
The form of \meff{} neglecting $m_1$ but 
including terms proportional to $s_3$ reads  
\be \label{eq:meffhie}
\meff \simeq \sqrt{\dms c_3^4 \; s_1^4 + \dma  s_3^4 + 2 \; 
\sqrt{\dms \dma}  s_1^2 \; s_3^2 \; c_3^2 \; c_{2\phi} } \, ,   
\ee
where $\phi = \alpha - \beta$. 
The maximal \meff{} for $t_3^2 = 0$ is 
about $7.7 \cdot 10^{-3}$ eV for a maximal \dms of about 
$3 \cdot 10^{-4}$ eV$^2$ in case of the LMA solution. For the other 
solutions it is around or below $10^{-4}$ eV\@. 
For the maximally allowed $t_3^2 = 0.08$ the effective mass can be 
at most $1.3 \cdot 10^{-2}$ ($5.5 \cdot 10^{-3}$, 
$5.2 \cdot 10^{-3}$) for LMA (LOW, VAC).  
For LMA and small $s_3$ the first term in Eq.\ (\ref{eq:meffhie}) 
dominates, for larger $s_3$ the third term can cancel the 
first two when $\phi \simeq \pi/2$. For $\alpha - \beta = \pi/4$ the 
solar and atmospheric contributions to \meff{} are just summed up. 
From values of \dms smaller than about $10^{-7}$ eV$^2$ on, 
the second term dominates in Eq.\ (\ref{eq:meffhie}), 
\meff{} is then proportional to $s_3^2$ and the dependence on $\phi$ 
practically vanishes. This can be seen in 
Fig.\ \ref{fig:meffhie}, where \meff{} is shown as a function of $\phi$  
for different $\dms\!\!$, $s_3^2$ and $t_1^2$. 
From (\ref{eq:meffhie}), one can infer the phase difference as 
\be \label{eq:meffhiealbe} 
\cos 2(\alpha - \beta) = \frac{\meff^2 - \dms s_1^4 - \dma s_3^4}
{2\, \sqrt{\dms \dma} s_1^2 \, s_3^2} \, .
\ee
As well known, it will be very difficult to measure \meff{} 
in the hierarchical scheme when LMA is not the solar solution, because 
in order to give an accessible \meff, $s_3^2$ has to be very 
close to its current limit. For the LMA solution, \meff{} can lie 
above $2 \cdot 10^{-3}$ eV even for vanishing $s_3$, which however 
requires large $t_1^2$ and \dms$\!\!$. If 
the two phases conspire to fulfill 
$\alpha \simeq \pi/2 + \beta$, then cancellation occurs and 
\meff{} can vanish. 
This happens also for the \mmp and \mpm configurations, i.e.\ 
when the second and third mass eigenstates have opposite signs. 
The \ppp and \mmp cases correspond to $\phi=0$.\\

Therefore, in case of the LMA solution and 
if $s_3^2$ is sizable (i.e.\ larger than 
about 0.03) and \meff{} lies below the GENIUS limit, then $\phi$ 
is located around $\pi/2$. The \ppp and \mmp configurations 
are then ruled out. Values of \meff{} considerably larger than 0.001 eV 
would show that $\phi$ is close to zero, which corresponds to the 
\ppp or \mmp signatures. For solutions with lower \dms and 
$t_3^2 \ls 0.03$ no 
statements can be made because the predicted \meff{} is below the GENIUS 
limit. In addition, there is practically no 
dependence on the phases, even for sizable $t_3^2$.

\subsection{\label{sec:meffinv}Inverse hierarchical scheme}
This scheme is realized if in the inverse scheme it holds 
$m_1 \simeq m_2 \gg m_3$. Thus, neglecting  
$m_3 s_3^2$, the effective mass reads  
\be \label{eq:meffinv}
\meff \simeq \sqrt{\dma} \sqrt{1 - 4 \, s_1^2 \, c_1^2 \, s_\alpha^2} 
= \frac{\sqrt{\dma}}{1 + t_1^2} 
\sqrt{(1 + t_1^2)^2 -  4 \; t_1^2 \; s_\alpha^2} 
\simeq \sqrt{\dma} c_\alpha\, , 
\ee
where the latter approximation holds for $t_1^2 \simeq 1$. 
Note that it is not a function of \dms$\!\!$. 
It allows for complete cancellation only for $t_1^2= 1$, i.e.\ in the 
LMA solution there should be a non--vanishing effective 
mass (larger than about $5 \cdot 10^{-3}$ eV), whereas 
LOW allows for complete cancellation. 
For the best--fit points given in the previous section 
the effective mass is predicted to be smaller than 0.07 eV, independent 
of the solar solution. 
Fig.\ \ref{fig:meffinv} shows \meff{} as a function of $\alpha$ 
for different $t_1^2$. One finds indeed very few 
dependence on $s_3^2$, $\beta$ and \dms$\!\!$. It is seen that for 
non--maximal solar mixing, \meff{} can be 
probed regardless of the phase. 
From Eq.\ (\ref{eq:meffinv}), $\alpha$ can be calculated for 
given $t_1^2$, \dma and \meff: 
\be \label{eq:meffinval}
s_\alpha^2 \simeq \frac{1}{4 \, t_1^2}(1 + t_1^2)^2  
\left(1 - \frac{\meff^2}{\dma}\right)
\simeq 1 - \frac{\meff^2}{\dma} \, , 
\ee
where the last approximation holds again for $t_1^2 \simeq 1$. 
If e.g.\ $\meff = 0.04 \, (0.03)$ eV, then one gets 
$s_\alpha^2 \simeq 0.4 \, (0.8)$ for the best--fit point of the LMA 
solution and $s_\alpha^2 \simeq 0.2 \, (0.6)$ for the LOW case. 
With $s_\alpha^2 \le 1$ it is possible to find the condition 
\be \label{eq:consinv}
\frac{\meff}{\sqrt{\dma}} \ge \frac{1 - t_1^2}{1 + t_1^2} \, ,  
\ee
under which the inverse hierarchical scheme is 
valid\footnote{In the following, we will use positive \meff. 
For $t_1^2 > 1$ it is understood that the absolute value of the 
right--hand side of (\ref{eq:consinv}) is taken.}. The number on 
the right--hand side lies 
between 1/9 and 2/3 for the LMA solution, between 
zero and 1/3 for the LOW case and zero and 1/2 for VAC\@. 
For instance, if $t_1^2 = 0.4 \; (0.7, 2.4)$, 
then $\meff \gs 0.02 \; (0.009, 0.02)$ eV\@. In this scenario, 
the best--fit points of the LMA and VAC solar solutions 
do not make any difference since they yield identical results for \meff.\\
 
Therefore, maximal solar mixing and \meff{} above 0.01 eV means that 
$\alpha$ is small or close to $\pi$, which corresponds also to the 
\ppp and \mmp cases. A value of \meff{} below the GENIUS limit 
implies that 
$\alpha \simeq \pi/2$, which is also possible for the 
\pmm and \mpm signatures. However, for such a small \meff{}  
a solar mixing angle very close to $\pi/4$ 
is required, i.e.\ if the LMA solution and the inverse 
scheme are verified, but \meff{} lies below the GENIUS bound, the inverse 
hierarchical scheme is ruled out. 
For non--maximal mixing, the dependence on 
$\alpha$ becomes smaller. Values of \meff{} 
below 0.01 eV are only possible for $\alpha \simeq \pi/2$ and 
$\theta_1 \simeq \pi/4 \pm 0.1$. 
Since in the inverse hierarchy the 
phase $\beta$ is connected with the smallest mass state $m_3$ as well as 
with the small quantity $s_3^2$, a determination of this parameter 
is very questionable. However, two of the four $CP$ conserving 
possibilities may be ruled out. The \ppp and \mmp cases as well as 
the \pmm and \mpm configurations give the same values 
because of the smallness of  $m_3 \, s_3^2$.

\subsection{\label{sec:meffdeg}Degenerate scheme}
This scheme is realized for 
$m_1^2 \simeq m_2^2 \simeq m_3^2 \equiv m_0^2 \gg \dma\!\!$. 
It is then useful to define an ``averaged mass'' 
$\tm = \meff/m_0$, which reads 
\be \label{eq:tmCPV}
\tm \equiv \frac{\meff}{m_0} = 
c_3^2 \, \sqrt{c_1^4 + s_1^4 + t_3^4 
+ 2 (s_1^2 \, t_3^2 \, c_{2(\alpha - \beta)}  
+ c_1^2 (s_1^2 \, c_{2 \alpha} + t_3^2 \, c_{2 \beta} ))} \; . 
\ee
No dependence on the solar $\Delta m^2$ exists. 
The four $CP$ conserving configurations can be written as 
\be \label{eq:tmCPC}
\tm = 
\left\{ 
\baz 
1                                & (+++) \\[0.2cm]
\frac{\D 1}{\D 1 + t_1^2} 
\left(1 - t_1^2 - 2 \, s_3^2\right)  & (+--) \\[0.2cm]
\frac{\D 1}{\D 1 + t_1^2} 
\left(1 - t_1^2 (1 - 2 \, s_3^2)\right)  & (-+-) \\[0.2cm]
     1 - 2 \, s_3^2             & (--+) 
\ea \right. . 
\ee
Note that the \ppp and \mmp cases have an averaged mass independent on the 
solar solution. For the \ppp configuration, $\meff = m_0$ and for the 
\mmp case $\meff = m_0 (1 - 2 s_3^2)$, which is identical to the result for  
\ppp for vanishing $\theta_3$. In the same limit, as well as for 
maximal solar mixing, the \pmm and \mpm cases are identical. 
In general, the \pmm and \mpm cases are connected via 
$\theta_1 \ra \pi/2 - \theta_1 $. Therefore, e.g.\ 
$t_1^2 = 0.5$ in the \pmm case is identical to 
$t_1^2 = 2.0$ in the \mpm configuration. This is however only interesting for 
the VAC solution, since the other ones have $t_1^2 \le 1$. 
If $t_1^2 < 1$, then the minimal \tm{} occurs for the 
\pmm configuration, if $t_1^2 > 1$, then in the \mpm case.\\
 
Cancellation can only occur for the \pmm and \mpm signatures, which 
however requires that $t_1^2$ is very close to one. In fact, 
for the \mpm case only $t_1^2 = 1$ together with $s_3^2 = 0$ 
can give full cancellation, whereas for \pmm also close to 
maximal solar mixing with non--vanishing $\theta_3$ is sufficient for 
complete cancellation. The minimal \tm{} for the \mmp signature 
is 0.84, in the LMA solution the range of \tm{} is 0.022 to 0.67 
(0.11 to 0.69) for the \pmm (\mpm$\!\!$) case, respectively. 
In the LOW solution, \tm{} ranges from zero to 1/3 (zero to 0.44) for 
\pmm (\mpm$\!\!$) and the VAC solution predicts \tm{} to lie 
between zero and 0.54 (zero and 1/2) for 
\pmm (\mpm$\!\!$).\\

We give a few examples for possible statements: 
for $t_1^2 < 1$ the minimal \meff{} is obtained for the 
\pmm signature, thus, if $\tm \ls 0.02$, then the LMA solution is ruled 
out. For $\tm \gs 0.7 \, (0.6)$, the \ppp or \mmp case has to 
be realized for the LMA (LOW, VAC) solution. For $\tm \gs 0.84$ the 
\pmm and \mpm cases are ruled out.\\

In analogy to Eq.\ (\ref{eq:meffinv}) one can in the degenerate scheme 
write an equation for \meff{} when $s_3^2$ is neglected: 
\be \label{eq:meffdeg}
\meff \simeq m_0 \sqrt{1 - 4 \, s_1^2 \, c_1^2 \, s_\alpha^2} 
= \frac{m_0}{1 + t_1^2} 
\sqrt{(1 + t_1^2)^2 -  4 \; t_1^2 \; s_\alpha^2} \simeq m_0 c_\alpha \, ,   
\ee
from which one obtains a formula for the phase $\alpha$, 
\be \label{eq:meffdegal}
s_\alpha^2 \simeq \frac{1}{4 \, t_1^2}(1 + t_1^2)^2  
\left(1 - \tm^2\right)
\simeq 1 - \tm^2 \, , 
\ee
where the last approximation holds again for $t_1^2 \simeq 1$. 
The corresponding equation (\ref{eq:meffinval}) 
for the inverse hierarchy should be a more appropriate 
relation since there the small quantity $s_3^2$ is multiplied with 
the smallest mass. In the degenerate scheme it contributes together 
with $m_0$, which for sizable $s_3^2$ could be a non--negligible number. 
In general, an area in $\alpha$--$\beta$ space can be identified, when 
a limit or value of $m_0$ or \meff{} is known \cite{phases,ichNPB}. 
The smaller \tm{} is, i.e.\ the more cancellation occurs, the closer 
$\alpha$ is to $\pi/2$. This however is equivalent to the 
\pmm and \mpm signatures. 
Since (\ref{eq:meffdeg}) allows cancellation only for $t_1^2 = 1$, 
a vanishing \meff{} in the LMA case 
for a neutrino mass of $m_0^2 \gg \dma\!\!$ would mean 
that $s_3^2$ is not zero. 
The consistency relation for the scenario in this section reads 
\be
\tm \ge \frac{1 - t_1^2}{1 + t_1^2} \, . 
\ee
Violation of this condition implies sizable $s_3^2$, which, 
from (\ref{eq:tmCPC}), can be obtained for $t_1^2 < 1$ as 
\be
s_3^2 = \frac{1}{2} \left(1 - t_1^2 - \tm (1 + t_1^2)\right) \, . 
\ee
For $t_1^2 > 1$ the \mpm case gives the minimal \meff, and 
$s_3^2$ is obtained in this case as 
\be
s_3^2 = \frac{1}{2 t_1^2} \left(\tm (1 + t_1^2) - (1 - t_1^2) \right) \, .
\ee\\

We finally comment on a small possibility to calculate the phase $\beta$ 
for the SMA solution. 
Since $s_1^2 \simeq 0$, \meff{} does hardly depend on the phase $\alpha$. 
If in Eq.\ (\ref{eq:tmCPV}) $c_1 \simeq 1$, then one gets  
\be \label{eq:SMAbe} 
s_\beta^2 \simeq \frac{\D 1 - \tm^2}{\D 4 \, s_3^2 \, c_3^2} \, . 
\ee
The condition under which this is possible can be obtained from 
$s_{\beta}^2 \le 1$ and reads  
\be
\tm \gs \sqrt{1 - 4 \, s_3^2 \, c_3^2 } \gs 0.84 \, .  
\ee
When \meff{} is close to 0.4 eV then 
this situation seems unlikely, 
since $s_3^2$ has to be close to its current limit and in addition 
$m_0$ must be close to the lowest experimentally accessible value in order 
to probe $\beta$.

\subsection{Partial hierarchical schemes}
These schemes are realized when the smallest mass state is 
of order of $\sqrt{\dma}$, say, between 0.01 and 0.1 eV\@. 

\subsubsection{Normal scheme}
In the ``normal partial hierarchical scheme'' 
one can define again an averaged mass $\meff/m_1$. In this scenario it 
can be obtained from Eq.\ (\ref{eq:tmCPV}) with the replacement 
\be
\frac{\meff}{\; m_1} = \tm\left(m_0 \rightarrow m_1, 
s_3^2 \rightarrow s_3^2 \sqrt{1 + \frac{\dma}{\D m_1^2}}\right) \, . 
\ee 
Again, no dependence on the solar $\Delta m^2$ is present. 
Depending on the value of $m_1$, the dependence on $t_3^2$ 
is more or less strengthened through the presence of the 
square root $w_n = \sqrt{1 + \frac{\dma}{\D m_1^2}}$. The maximal 
\meff{} is given by $\meff \! \simeq m_1 (1 + s_3^2 \, w_n) \ls 0.11$ eV\@. 
For vanishing $t_3^2$ complete cancellation is again only 
possible for $t_1^2 = 1$, i.e.\ not for the LMA solution. 
The situation is then equivalent to the one for the inverse hierarchical 
scheme. For vanishing $t_3^2$ one can write an equation for 
$\alpha$ in analogy to (\ref{eq:meffinval}), with the 
replacement $\dma \!\! \rightarrow m_1^2$.\\
 
For $t_1^2 = 0.4$, \meff{} lies below 0.01 eV only for $\alpha \simeq \pi/2$ 
and $m_1 \ls 0.02$ eV\@. If also $\beta \simeq \pi/2$ and 
$t_3^2 \gs 0.02$, then \meff{} can lie below the GENIUS bound. 
With increasing $t_1^2$, the minimal 
value of $m_1$, for which this happens, is increasing. 
Values below $2 \cdot 10^{-3}$ eV are only possible if 
$m_1$ and $t_1^2$ are small and $t_3^2$ is sizable or if 
the mixing is close to maximal and $t_3^2$ is small. 
When $t_1^2 \gs 1.5$, similar statements hold, however, $\beta \simeq 0$ 
is now required in order to allow for large cancellations. 
Thus, for large $t_1^2 < 1$ ($t_1^2 > 1$), sizable $t_3^2$ and 
$\meff < 2 \cdot 10^{-3}$ eV, then $m_1$ has to be small, 
$\alpha$ around $\pi/2$ and 
$\beta$ around $\pi/2$ ($\pi$). This situation corresponds to the 
\pmm ($\mpm \!\!$) configuration. 
For close to maximal solar mixing and $\meff < 2 \cdot 10^{-3}$ eV, 
$\alpha$ is around $\pi/2$. 
If $t_3^2$ is sizable, then in addition $m_1$ has to be small. 
When \meff{} is around 0.01 eV, the phase $\alpha$ has to be small, 
which corresponds to the \pmm or \mpm signature.

\subsubsection{Inverse scheme}
In the ``inverse partial hierarchical scheme'' $\meff/m_3$ 
can be obtained from 
Eq.\ (\ref{eq:tmCPV}) with the replacement 
\be
\frac{\meff}{\; m_3} = \tm\left(m_0 \rightarrow m_3, 
c_1^2 \rightarrow c_1^2 \sqrt{1 + \frac{\dma}{\D m_3^2}}, 
s_1^2 \rightarrow s_1^2 \sqrt{1 + \frac{\dma}{\D m_3^2}}
\right) \, , 
\ee
which has a slightly weaker dependence on $t_3^2$ than in the normal 
hierarchy, because the contributions of $c_1^2$ and $s_1^2$ 
are enhanced by the 
factor $w_i = \sqrt{1 + \frac{\dma}{\D m_3^2}}$. The maximal \meff{} 
is given by $\meff \simeq m_3 \, (w_i + s_3^2) \ls 0.12$ eV\@. For vanishing 
$t_3^2$ one can obtain $\alpha$ via Eq.\ (\ref{eq:meffinval}) and 
the replacement $\dma \!\! \rightarrow w_i^2 \, m_3^2 = m_3^2 + \dma\!\!$. 
For $t_1^2 \ls 0.6$ and $t_1^2 \gs 1.5$, 
\meff{} lies always above 0.01 eV\@. For closer to maximal 
(but not exactly maximal) 
mixing it can lie below 0.01 eV for large $t_3^2$ and low $m_3$ or only for 
small $t_3^2$ as long as $\alpha \simeq \pi/2$ and $\beta \simeq \pi/2$ (0) 
for $t_1^2 < 1$ ($t_1^2 > 1$). 
Thus, values below 0.01 eV are possible when the solar mixing is maximal 
and $\alpha \simeq \pi/2$ (which is equivalent to the \pmm or \mpm signature) 
or if for $t_1^2 < 1$ the second phase $\beta$ is around $\pi/2$, which 
is the \pmm configuration. For $t_1^2 > 1$, $\beta$ should be close to 
$\pi$, which corresponds to the \mpm case.\\

Once we finished now the discussion of the special hierarchies, 
we can order them with respect to the maximal \meff{} they predict: 
\be
\mbox{degenerate} > \mbox{partial inverse} > \mbox{partial normal} > 
\mbox{inverse} > \mbox{normal} \; . 
\ee

\section{\label{sec:gen}General case}
We shall use the best--fit 
oscillation parameters as given in Eqs.\ (\ref{eq:atmBF},\ref{eq:solBF}) 
and assume the following uncertainties of the solar $\Delta m^2$: 
5 $\%$ for LMA, 10 $\%$ for LOW and 5 $\permil$ for VAC\@. 
For $\tan^2 \theta_1$ and \dma we assume an uncertainty of 5 and 10 $\%$, 
respectively. The effective mass is analyzed as a function of the 
smallest mass state for different $t_3^2$. What results with these 
assumptions is an area in parameter space, which denotes 
the region between the maximal and minimal \meff. 
Unless otherwise 
stated, the area for the \ppp case is so small that it appears as a line.

\subsection{Normal scheme}  
For the normal hierarchy, the result is shown in 
Figs.\ \ref{fig:LMAn1} to \ref{fig:VACn1}. The structure of the 
``$CP$--violating'' area is the less complicated the smaller 
$t_3^2$ is. From $t_3^2$ smaller than about $10^{-3}$ ($10^{-4}$)  
on, the \mpm and \pmm signatures become indistinguishable for the LMA 
(LOW, VAC) solution. Up to six separated areas exist, 
for $t_3^2 \ls 10^{-3}$ they merge into one, which area is smaller than 
the sum of the areas for sizable $t_3^2$. 
For the LOW solution the area is significantly larger than for LMA and for 
VAC\@. In case of VAC the area is smallest.\\

In Fig.\ \ref{fig:un} the consequences of different uncertainties 
of the oscillation parameters are shown. We concentrate on the 
LMA solution, $t_3^2 = 0.01$ and start with a typical allowed 
parameter space, which is 
$\dma \!\! = (1.5 \ldots 4) \cdot 10^{-3}$ eV$^2$, 
$\dms \!\! = (1 \ldots 12) \cdot 10^{-5}$ eV$^2$ and 
$t_1^2 = 0.2 \ldots 0.8$, which is denoted as ``everything''. 
Then it is allowed for uncertainties of the solar and atmospheric 
parameters around the best--fit values of 
Eqs.\ (\ref{eq:atmBF},\ref{eq:solBF}), which 
are indicated in the figure. 
The lower right plot 
is for exact measurements, which coincides with the situation 
analyzed in \cite{petwol}. Similar plots have been presented first 
in \cite{petpas}, where the $\Delta m^2$ have been allowed to vary within 
their $90 \%$ C.L. values and different $t_1^2$ and $t_3^2$ have been 
taken. The situation under study in the present paper is more 
accurate with respect to the expected future uncertainty of the 
oscillation parameters. Now an area for the \ppp case can be 
identified. For an uncertainty of $5 \%$ and $10 \%$ for the solar 
and atmospheric parameters (as used in Fig.\ \ref{fig:LMAn1}), the area 
becomes again a line and consequently 
is not shown anymore. Currently, there is 
only a small $CP$--violating area, between the minimal \mmp and 
the maximal \mpm line,  
although it exists at large \meff{} and $m_1$. 
The area grows with decreasing 
uncertainty and takes the complicated form known from 
the previous figures when the solar parameters are known 
to a precision better than 10 $\%$. 
These additional areas appear however around or below the 
maximal GENIUS limit.

\subsection{Inverse scheme}
The plots with the $CP$--violating areas are shown in 
Figs.\ \ref{fig:LMAi1}, \ref{fig:LOWi1} and  \ref{fig:VACi1} for the LMA, 
LOW and VAC solution. The situation is now much simpler, $CP$ violation 
occurs only between the minimal \mmp and the maximal \mpm line. 
There is also a small area between the minimal \mpm and the 
maximal \pmm line, which disappears for $t_3^2 \ls 0.01$. 
The \pmm and \mpm signatures become indistinguishable 
for $t_3^2 \ls 0.01$.
With the chosen oscillation parameters there 
is no complete cancellation and therefore all expected \meff{} values 
lie above 0.01 (0.006) eV for the LMA and VAC (LOW) case. Again, 
the LOW case provides the largest $CP$--violating area, the LMA 
and VAC areas are of comparable size.\\

As for the normal scheme, large part of the areas cover a range of \meff{} 
that is larger than the expected $20 \%$ uncertainty of the experimental 
results. This is negligible with respect to the uncertainty stemming 
from the NME calculations. Consequently, these have 
to be overcome in order to make reasonable statements (not only) 
about the presence of $CP$ violation in \obb.

\section{\label{sec:sum}Final remarks and summary}
The presence of leptonic $CP$ 
violation especially in \obb{} can strengthen our believe in leptogenesis, 
the creation of a baryon asymmetry through 
out--of--equilibrium decays of heavy 
Majorana neutrinos. These heavy neutrinos are also responsible 
for the light neutrino masses through the see--saw mechanism, 
linking thus neutrino oscillations with leptogenesis.  
Baryon number and 
$CP$ violation are necessary conditions for creating a baryon asymmetry. 
Given that in most models the heavy Majorana neutrinos are too heavy 
($\gs 10^{10}$ GeV) to 
be produced at realistic collider energies, the demonstration of 
lepton number violation and leptonic $CP$ violation could be 
the only possibility to validate leptogenesis. 
A general feature of models presented in the literature 
is the dependence of the 
baryon asymmetry $Y_B$ on the Majorana phases $\alpha$ and $\beta$. 
For example, in the left--right symmetric model presented in 
\cite{wir}, a sufficient $Y_B$ can be 
generated without the ``Dirac phase'' $\delta$.  
This has also been observed in the minimal $SO(10)$ 
model analyzed in \cite{branco}. The presence of $CP$ violation in 
\obb{} is required there to produce the correct amount of $Y_B$. 
This is why $CP$ violation in \obb{} plays an important role.\\

In the light of recent data we analyzed the presence of $CP$ 
violation in neutrinoless double beta decay. The observed non--maximality 
of the solar mixing in case of the LMA solution allowed to make 
some statements about possible cancellations. 
The four $CP$ conserving sign signatures can in many cases 
be grouped into two pairs, in some cases even one unique 
solution can be identified. 
In the hierarchical scheme the \ppp and \pmm cases are 
equivalent because \meff{} depends on the difference of two phases. 
In the inverse hierarchical scheme only one phase can be probed, 
which leads to identical results for the 
\ppp and \mmp cases as well as for the \pmm and \mpm cases. 
Due to the large solar mixing and the smallness of $s_3^2$, these 
two pairs also exist for the degenerate and partial hierarchical schemes. 
Simple formulas for the Majorana phases and 
consistency relations for these hierarchies 
have been collected and the different situation for values of $t_1^2$ 
smaller or bigger than one has been commented on. 
The $CP$ violating areas including present and future uncertainties  
of the mixing parameters were identified. The LOW solution provides 
the best opportunity to establish the presence of leptonic $CP$ 
violation, since the relevant area in parameter 
space is largest in this case, regardless of the mass scheme. 
Obviously, in the inverse scheme, where for small neutrino masses 
the predicted \meff{} is considerably higher, the situation is better. 
However, the uncertainty stemming from the calculation of the 
nuclear matrix elements will remain the big drawback for this 
possibility.

\hspace{3cm}
\begin{center}
{\bf \large Acknowledgments}
\end{center}
I thank S.T.~Petcov for helpful comments. 
This work has been supported in part by the
``Bundesministerium f\"ur Bildung, Wissenschaft, Forschung und
Technologie'', Bonn under contract No. 05HT9PEA5.

\begin{center}
\begin{figure}[hp]
\begin{center}
\epsfig{file=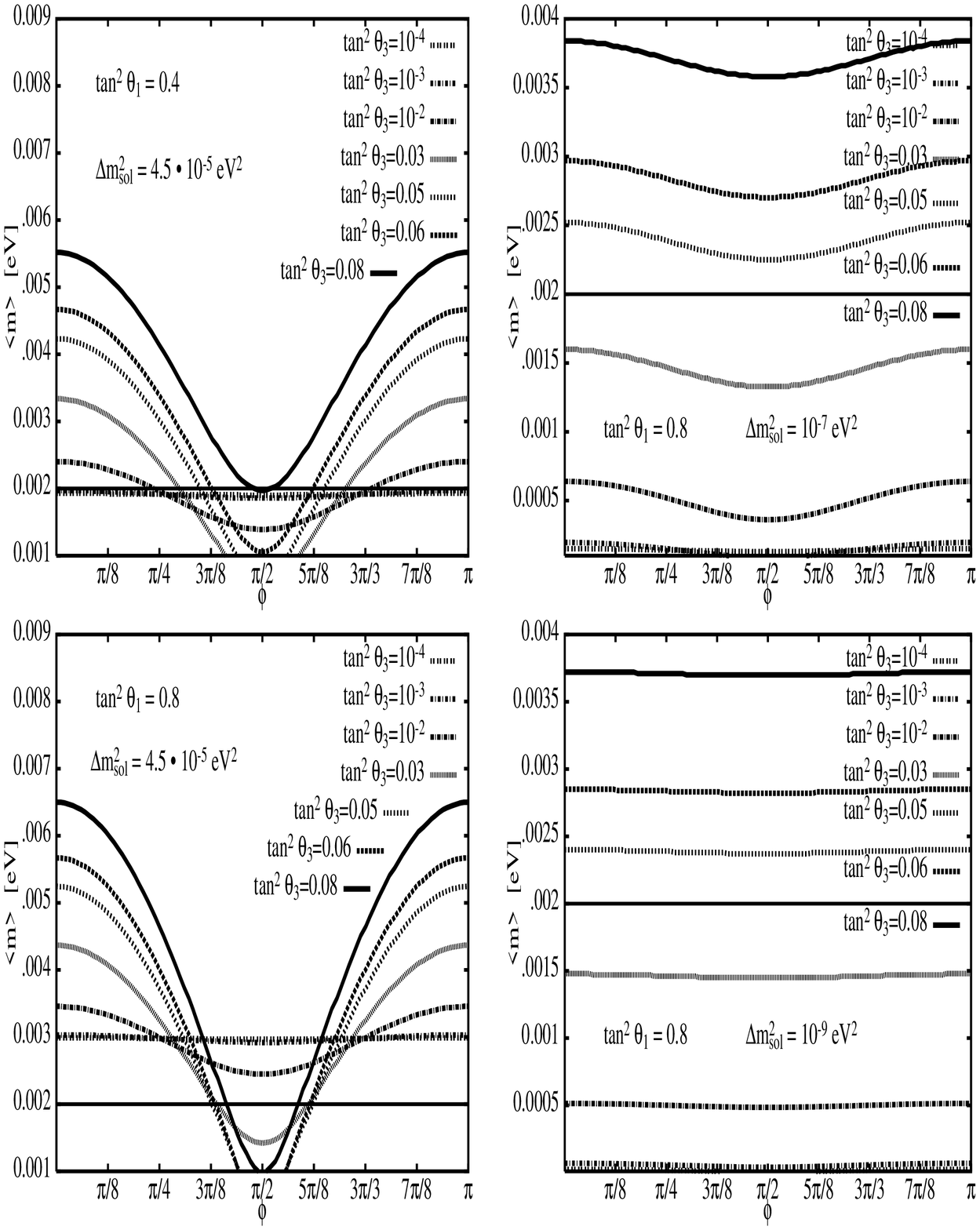,width=17cm,height=22cm}
\end{center}
\vspace{-2.cm}
\caption{\label{fig:meffhie}The effective mass \meff{} in the 
hierarchical scheme 
as a function of $\phi = \alpha - \beta$ for different $\tan^2 \theta_3$, 
$\tan^2 \theta_1$ and \dms$\!\!$. 
The smallest mass is $m_1 = 10^{-5}$ eV\@, 
$\dma \!\! = 2.5 \cdot 10^{-3} \, {\rm eV}^2$ 
and $\beta$ is chosen to be zero.}
\end{figure}

\begin{figure}[hp]
\begin{center}
\epsfig{file=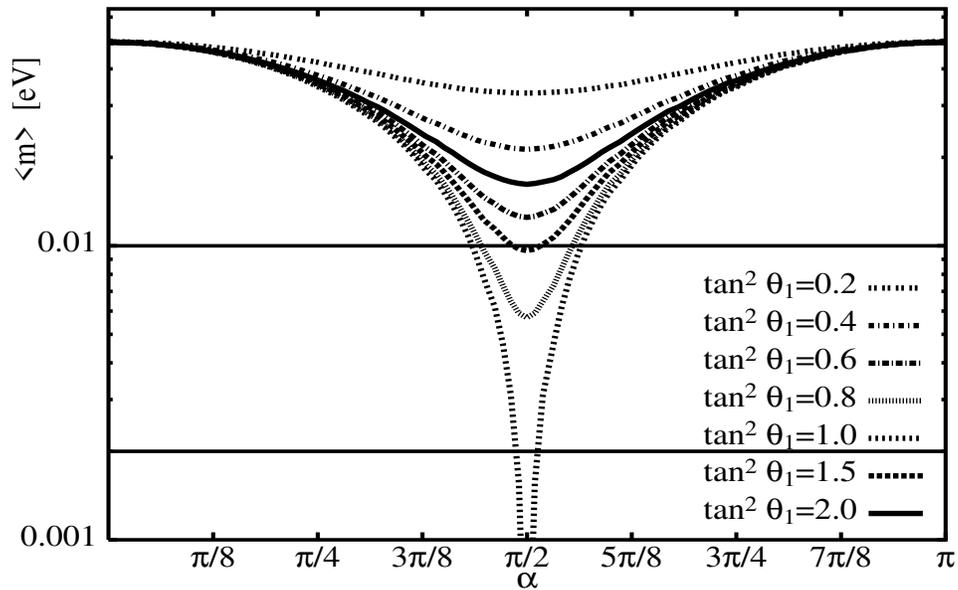,width=13cm,height=8cm}
\end{center}
\caption{\label{fig:meffinv}The effective mass \meff{} in the 
inverse hierarchical scheme 
as a function of $\alpha$ for different $t_1^2$. 
The smallest mass is $m_3 = 10^{-5}$ eV\@, 
$\dma \!\! = 2.5 \cdot 10^{-3} \, {\rm eV}^2$, 
$\dms \!\! = 4.5 \cdot 10^{-5} \, {\rm eV}^2$, 
$s_3^2 = 0.01$ and $\beta = 0$.} 
\end{figure}  
\end{center}

\pagestyle{empty}

\begin{center}
\begin{figure}[hp]
\vspace{-3cm}
\hspace{-20mm}
\epsfig{file=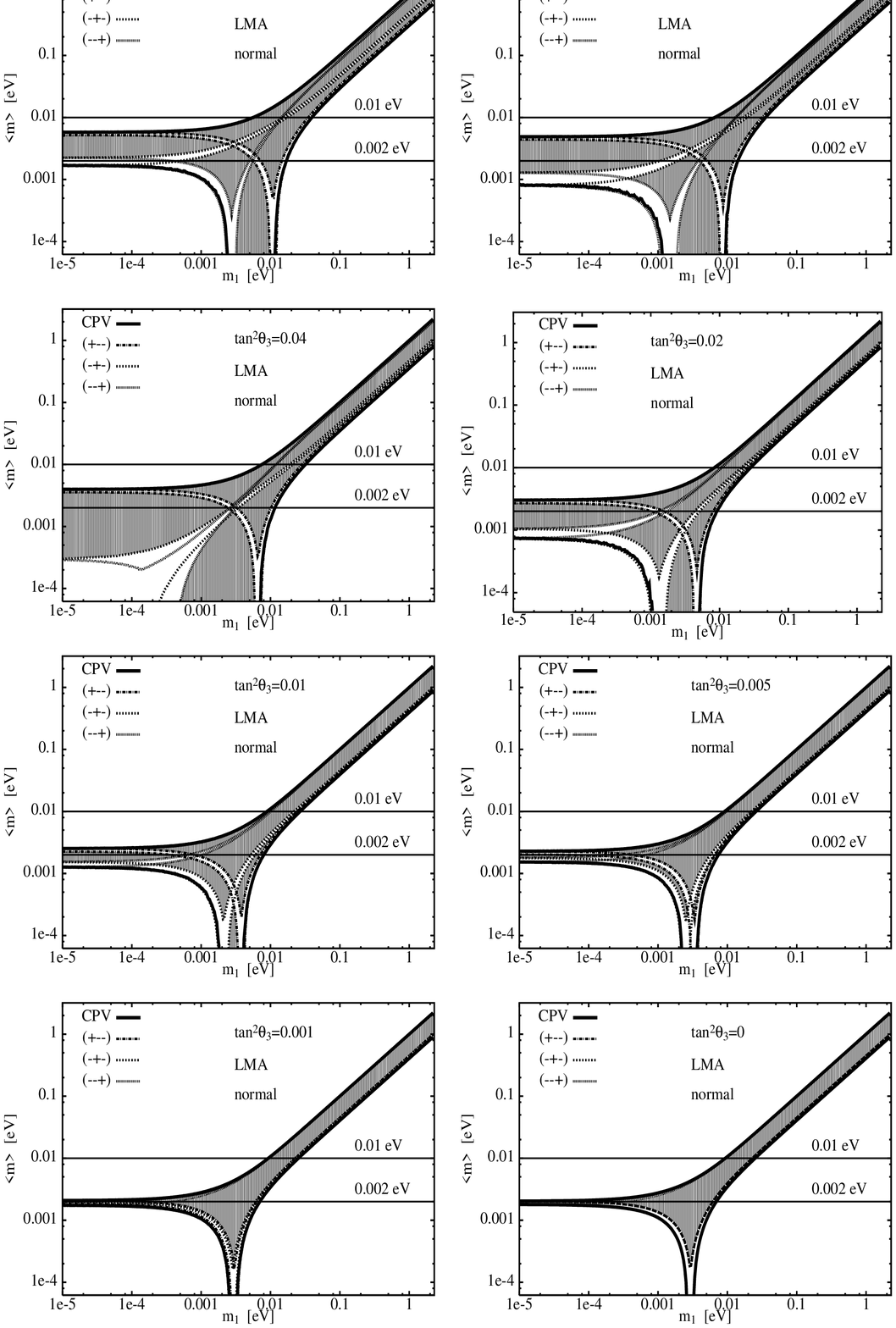,width=21cm,height=24cm}
\caption{\label{fig:LMAn1}The range of \meff{} in the normal scheme 
for the LMA solution and an uncertainty of the oscillation parameters 
as described in the text. 
The ``$CP$ violating'' area is indicated by the hatched area.}
\end{figure}

\begin{figure}[hp]
\vspace{-3cm}
\hspace{-20mm}
\epsfig{file=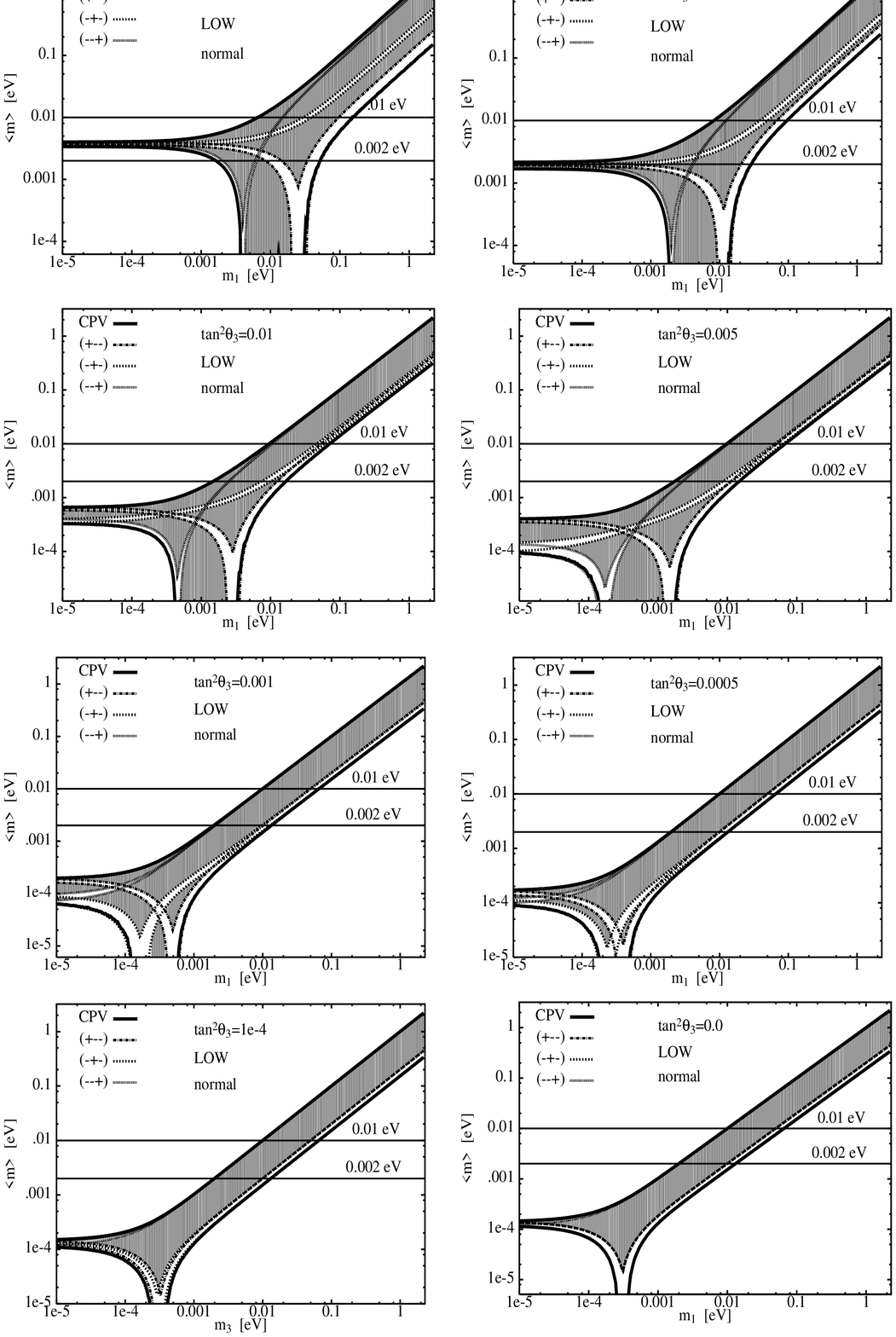,width=21cm,height=24cm}
\caption{\label{fig:LOWn1}The range of \meff{} in the normal scheme 
for the LOW solution and an uncertainty of the oscillation parameters 
as described in the text. 
The ``$CP$ violating'' area is indicated by the hatched area.}
\end{figure}

\begin{figure}[hp]
\vspace{-3cm}
\hspace{-20mm}
\epsfig{file=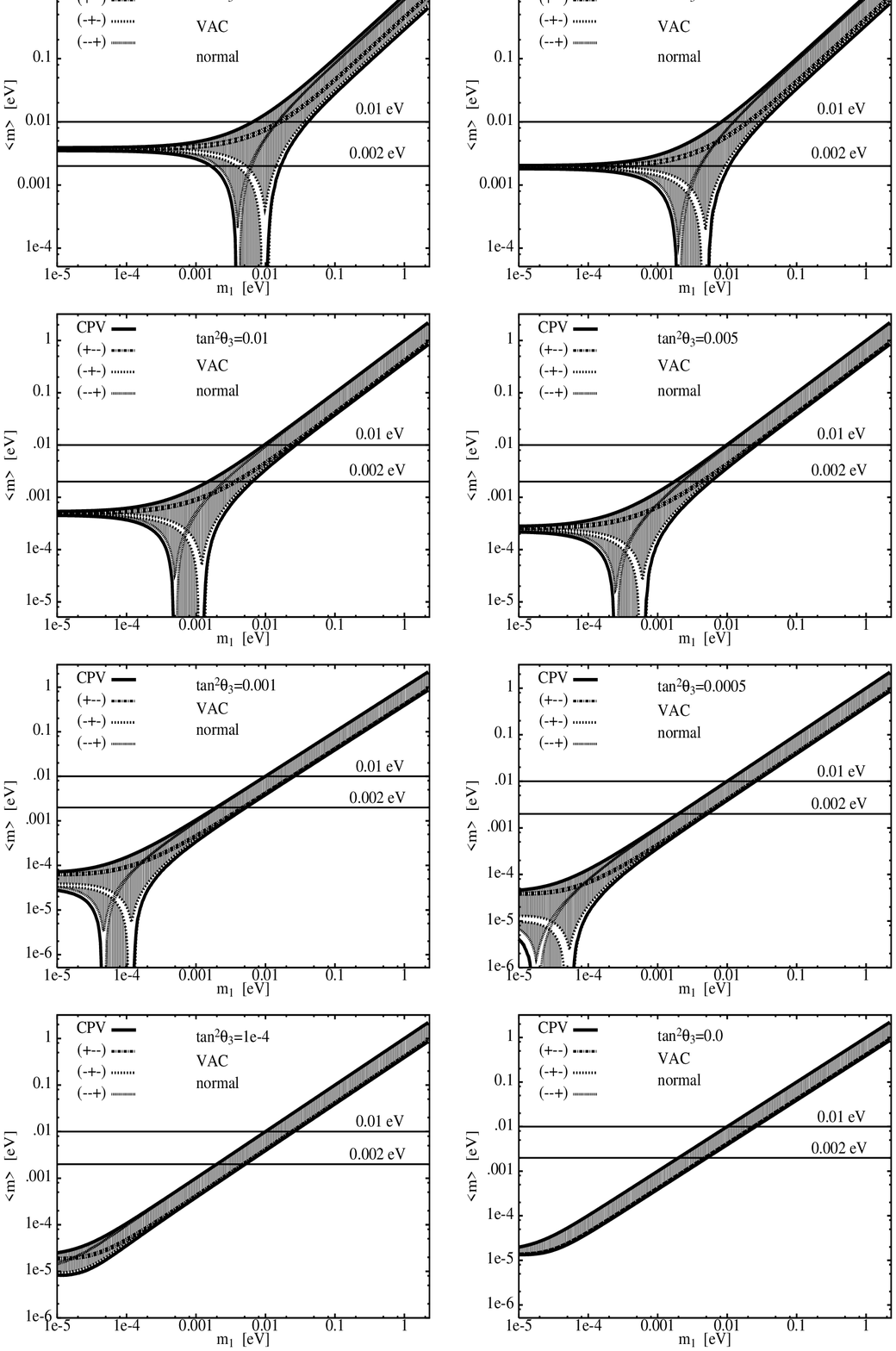,width=21cm,height=24cm}
\caption{\label{fig:VACn1}The range of \meff{} in the normal scheme 
for the VAC solution and an uncertainty of the oscillation parameters 
as described in the text. 
The ``$CP$ violating'' area is indicated by the hatched area.}
\end{figure}

\begin{figure}[hp]
\vspace{-3cm}
\hspace{-20mm}
\epsfig{file=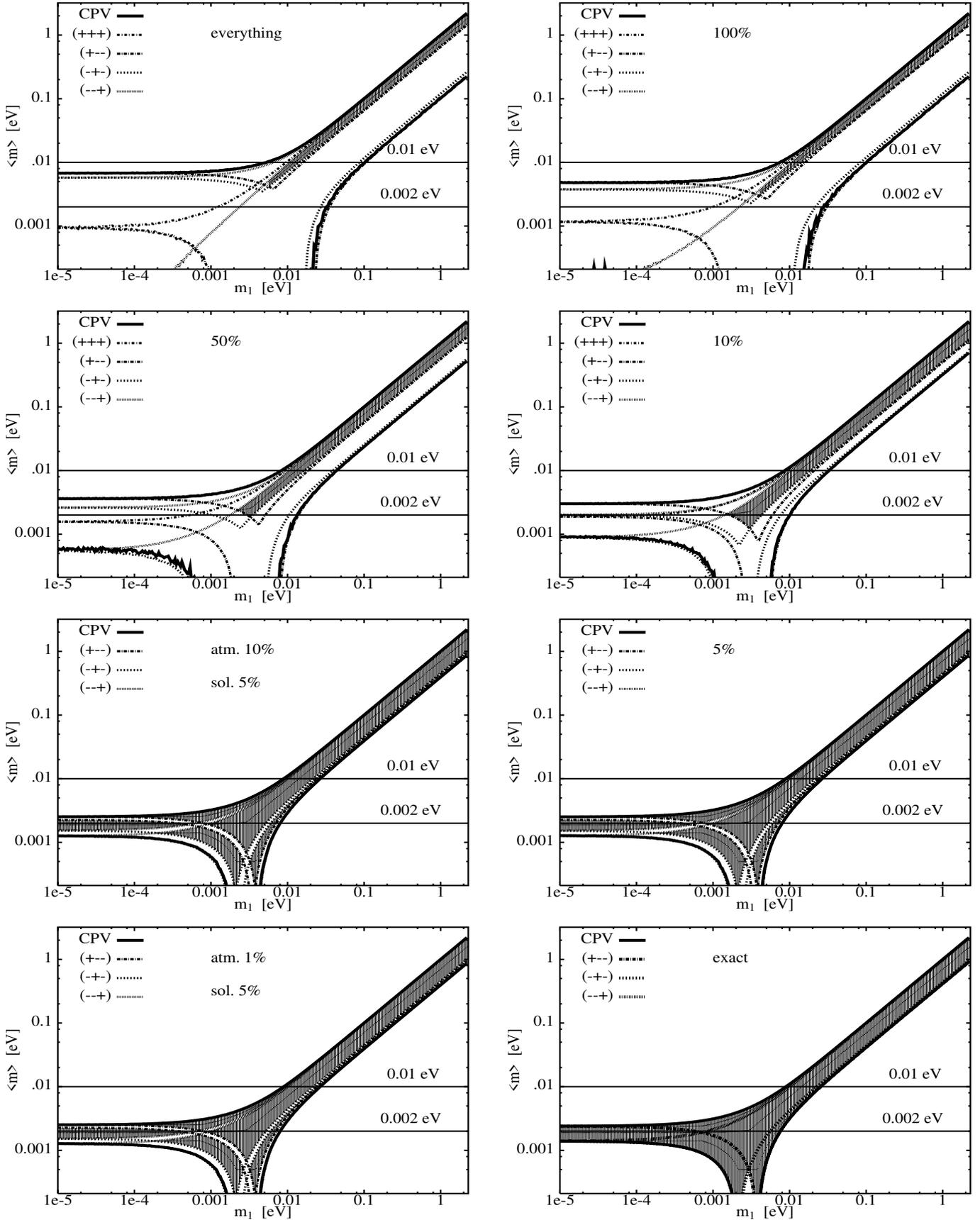,width=21cm,height=24cm}
\caption{\label{fig:un}The range of \meff{} in the normal scheme 
for the LMA solution, $t_3^2 = 0.01$ and different 
uncertainties of the oscillation parameters 
as described in the text. 
The ``$CP$ violating'' area is indicated by the hatched area.}
\end{figure}

\begin{figure}[hp]
\vspace{-3cm}
\hspace{-20mm}
\epsfig{file=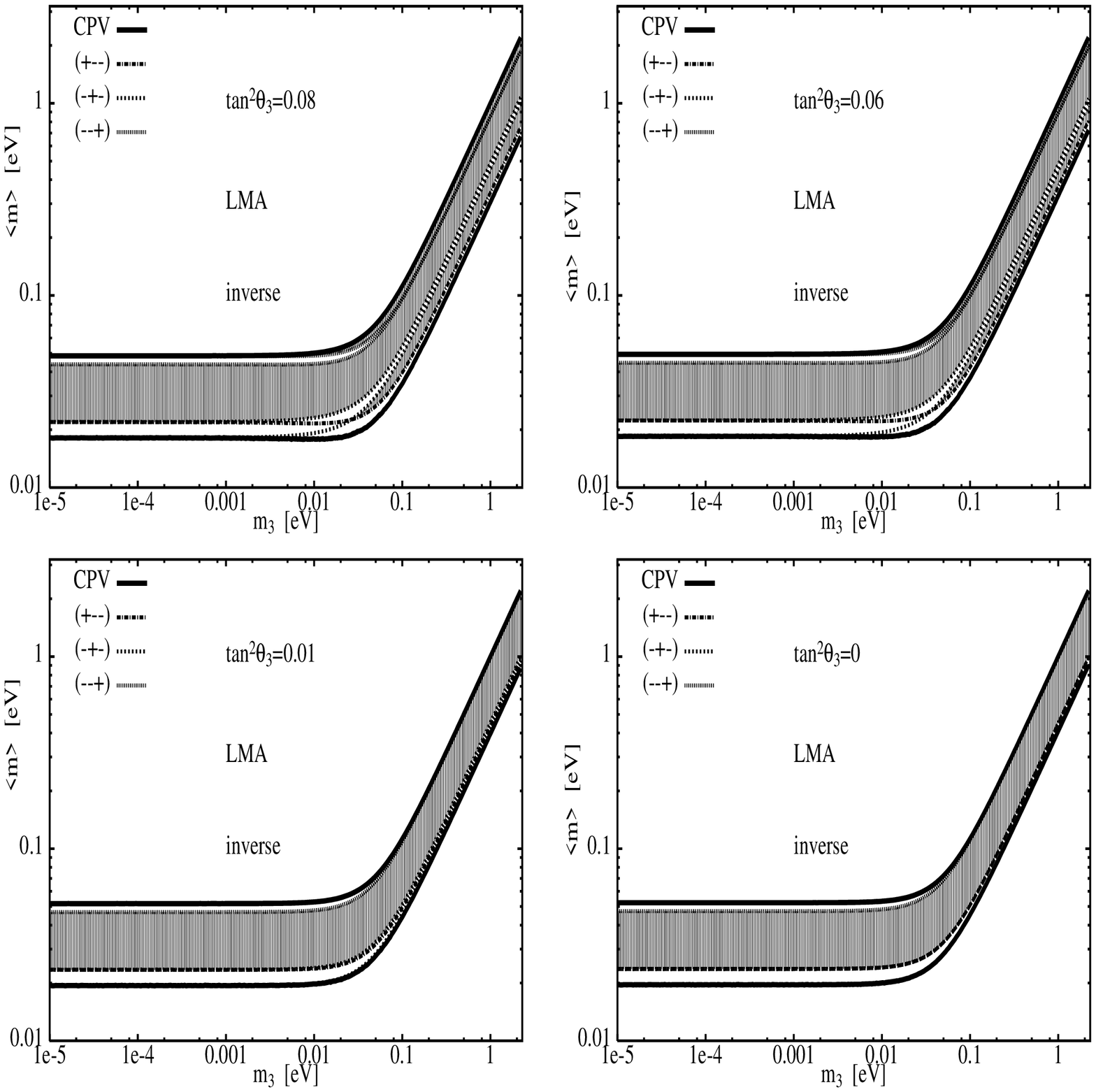,width=18cm,height=25cm}
\vspace{-5cm}
\caption{\label{fig:LMAi1}The range of \meff{} in the inverse scheme 
for the LMA solution and an uncertainty of the oscillation parameters 
as described in the text. 
The ``$CP$ violating'' area is indicated by the hatched area.}
\end{figure}

\begin{figure}[hp]
\vspace{-3cm}
\hspace{-20mm}
\epsfig{file=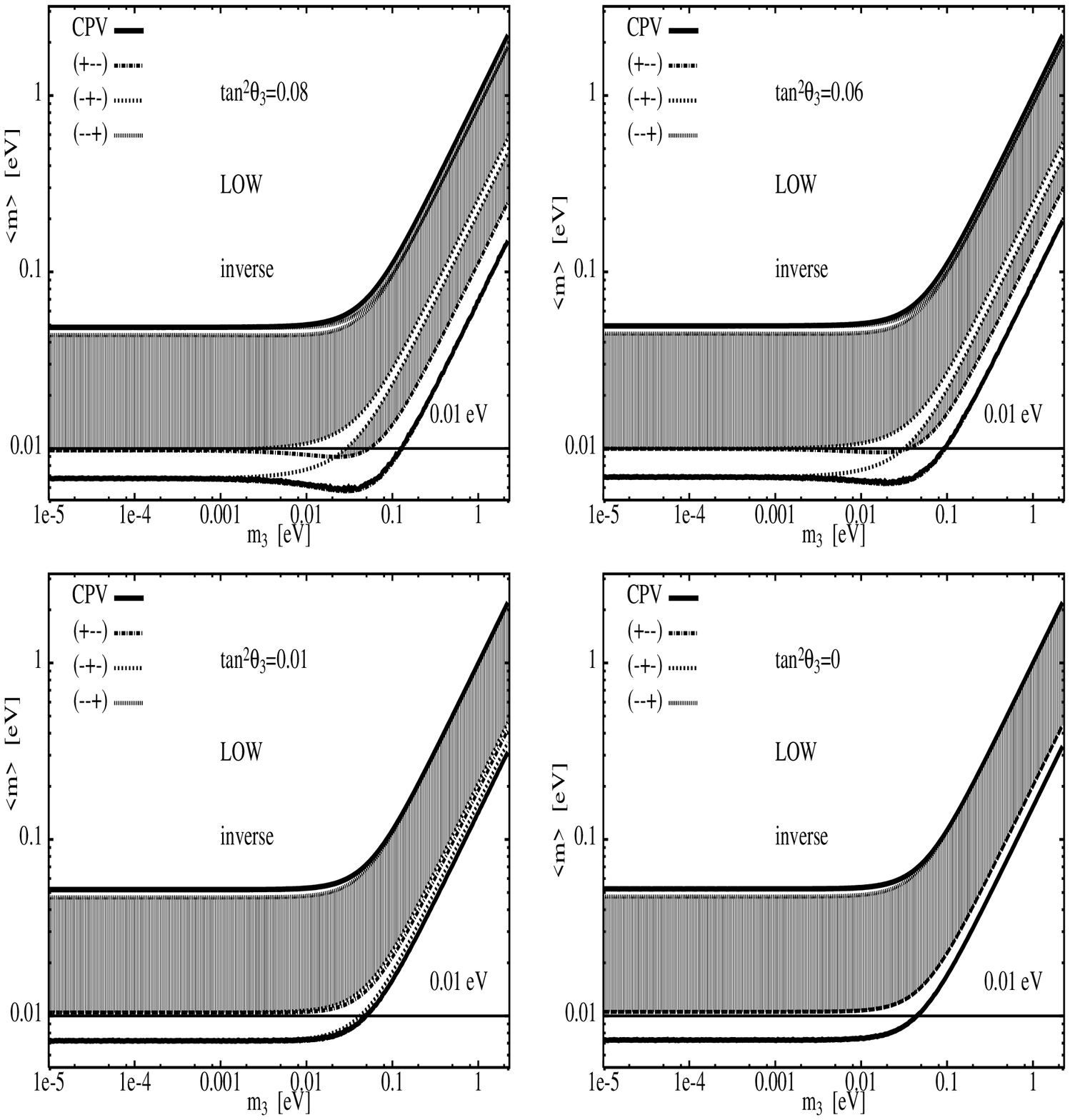,width=18cm,height=25cm}
\vspace{-5cm}
\caption{\label{fig:LOWi1}The range of \meff{} in the inverse scheme 
for the LOW solution and an uncertainty of the oscillation parameters 
as described in the text. 
The ``$CP$ violating'' area is indicated by the hatched area.}
\end{figure}

\begin{figure}[hp]
\vspace{-3cm}
\hspace{-20mm}
\epsfig{file=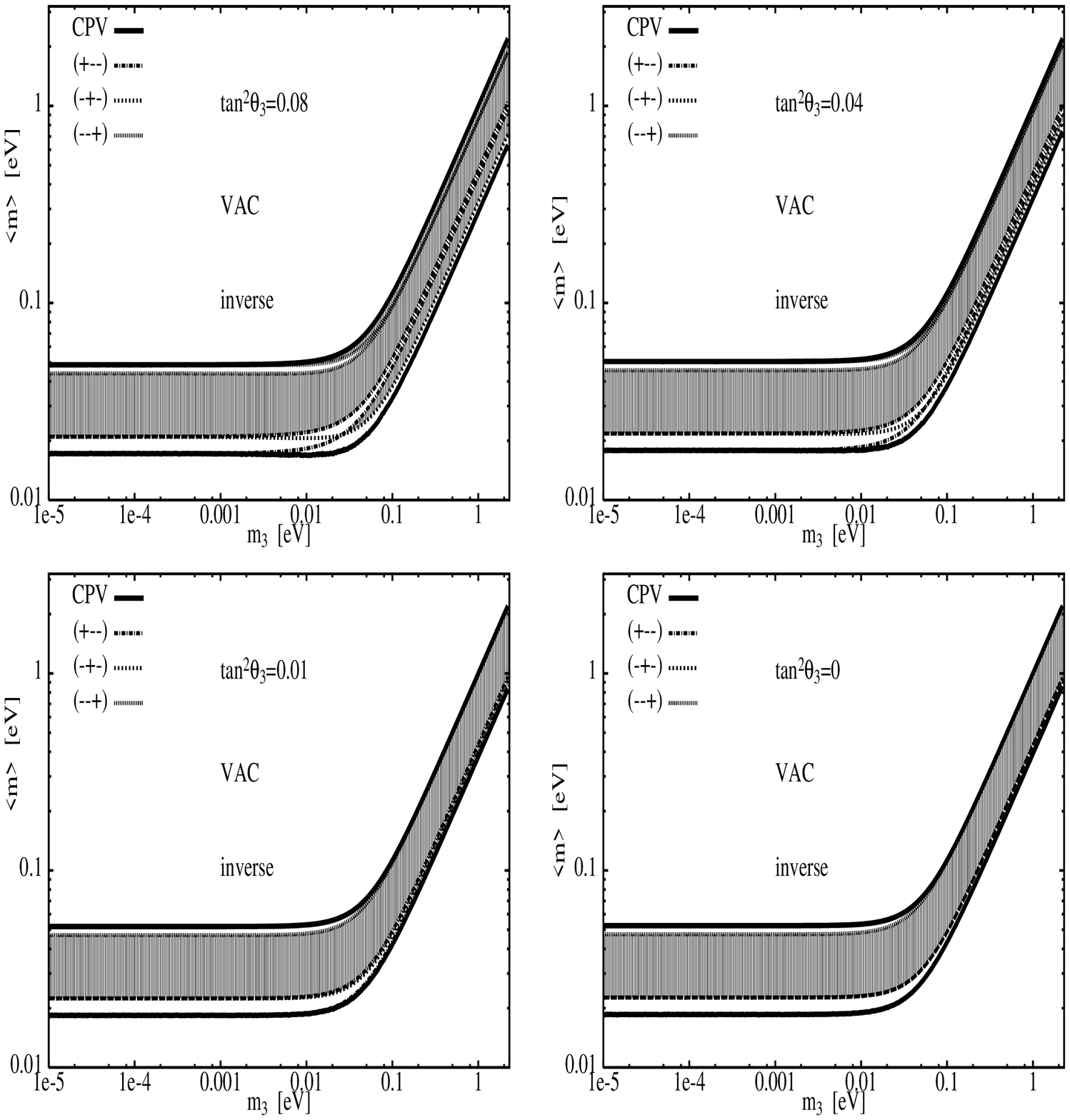,width=18cm,height=25cm}
\vspace{-5cm}
\caption{\label{fig:VACi1}The range of \meff{} in the inverse scheme 
for the VAC solution and an uncertainty of the oscillation parameters 
as described in the text. 
The ``$CP$ violating'' area is indicated by the hatched area.}
\end{figure}

\end{center}

\end{document}